# Title: Freeform Diffractive Metagrating Design Based on Generative Adversarial Networks

**Authors:**

Jiaqi Jiang[1], David Sell[2], Stephan Hoyer[3], Jason Hickey[3], Jianji Yang[1], and Jonathan A. Fan[1]

**Affiliations:**

[1]Department of Electrical Engineering, Stanford University, Stanford, CA, USA.

[2]Department of Applied Physics, Stanford University, Stanford, CA, USA.

[3]Google AI Applied Science, Mountain View, CA, USA.

**Abstract:**

A key challenge in metasurface design is the development of algorithms that can effectively and efficiently produce high performance devices. Design methods based on iterative optimization can push the performance limits of metasurfaces, but they require extensive computational resources that limit their implementation to small numbers of microscale devices. We show that generative neural networks can train from images of periodic, topology-optimized metagratings to produce high-efficiency, topologically complex devices operating over a broad range of deflection angles and wavelengths. Further iterative optimization of these designs yields devices with enhanced robustness and efficiencies, and these devices can be utilized as additional training data for network refinement. In this manner, generative networks can be trained, with a onetime computation cost, and used as a design tool to facilitate the production of near-optimal, topologically-complex device designs. We envision that such data-driven design methodologies can apply to other physical sciences domains that require the design of functional elements operating across a wide parameter space.

Metasurfaces are foundational devices for wavefront engineering *(1-3)*. They can focus and steer an incident wave *(4)* and manipulate its polarization *(5)* in nearly arbitrary ways, surpassing the limits set by conventional optics. They can also shape and filter spectral features, which has practical applications in sensing *(6)*. In addition, metasurfaces have been implemented in frameworks as diverse as holography *(7)* and transformation optics *(8)*, and they can be used to perform mathematical operations with light *(9)*.

Iterative optimization methods, including adjoint-based *(10)* and objective-first *(11)* topology optimization, are effective and general techniques for producing high performance metasurface designs *(12-16)*. Devices designed with these principles have curvilinear freeform layouts, and they utilize non-intuitive optical interactions to achieve high efficiencies *(21)*. A principle challenge with iterative optimizers is that they require immense computational resources, limiting their application to small numbers of microscale devices. This is problematic for metasurfaces, where large ensembles of microscopic devices operating with differing input and output angles, output phases, wavelengths, device materials, thicknesses, and polarizations, amongst other parameters, are desired to construct macroscale diffractive elements *(17, 34)*. In this context, it would be of immense value for optical engineers to have access to a computational design tool capable of expediting the realization of high performance metasurface designs, given a desired set of operating parameters. Such a tool would also enable the generation of large device datasets that can combine with data mining analyses to unveil the underlying physics and design principles behind topology-optimized devices.

In this Article, we present a metasurface design platform combining conditional generative adversarial networks (GANs) *(18, 19)* with iterative optimization that serves as an effective and computationally efficient tool to produce high performance metasurfaces. GANs are deep generative neural networks originating from the computer vision community, and they are capable of learning geometric features from a set of training images and then generating new images based on these features. They have been explored previously in the design of subwavelength-scale optical nanostructures *(39),* and we investigate their potential in optimizing high performance diffractive optical devices. As a model system, we will focus our study on silicon metagratings *(20,21)* that deflect electromagnetic waves to the +1 diffraction order. An outline of our design platform is presented in Figure 1. To design metagratings with a desired set of outgoing angles and operating wavelengths, we train a conditional GAN on a training set of high efficiency device images and then generate many candidate device images with a diversity of geometric shapes. These devices are then characterized using a high-speed electromagnetics simulator, and the corresponding high efficiency devices are further refined using iterative optimization. These final metagrating layouts serve as new training data to retrain the conditional GAN and expand its overall capabilities.

**Results and discussion**

The starting point is the production of a high-quality training set consisting of 600 high-resolution images of topology-optimized metagratings, for GAN training (Figure 2A and Figure 2B). This initial training set is orders of magnitude smaller than those used in conventional machine vision applications *(22)*. Each device is 325 nm-thick and designed to operate at a wavelength between 800 nm and 1000 nm, in increments of 20 nm, and at an angle between 55 and 65 degrees, in increments of 5 degrees. For each wavelength and angle pair, we generate a distribution of devices with a range of efficiencies, using different random dielectric starting layouts. We then keep the devices operating in the top 40$^{th}$ percentile of the efficiency distribution, which we term "above threshold" devices (see Figure S1 for distribution examples). We found that if we do not filter for "above threshold" devices, our GAN performs worse (Figure S2), indicating the need for exclusively high efficiency devices for training. An analysis of a GAN trained with devices possessing sparsely distributed wavelength and angle values is summarized in Figure S3 and shows comparable results to those here.

These data are used to train our conditional GAN, which consists of two deep networks, a generator and a discriminator (Figure 2C). The generator is conditioned to produce images of new devices as a function of deflection angle and operating wavelength. Its inputs are the metagrating deflection angle, operating wavelength, and an array of normally-distributed random numbers, which provides diversity to the generated device layouts. The discriminator helps to train the generator on what images to create by learning what constitutes a high performance device. Specifically, the discriminator trains to distinguish between actual devices from the training set and those from the generator.

The training process, in which the generator and discriminator are trained in alternating steps, can be described as a two-player game in which the generator tries to fool the discriminator by generating realistic-looking devices, while the discriminator tries to identify and reject generated devices from a pool of generated and real devices. Upon training completion, the discriminator will be able to identify the small differences between the generated and actual devices, while the generator will have learned how to produce images that could fool the discriminator. In other words, the generator will have learned the underlying topological features from optimized metagratings and be able to produce new, topologically complex devices for a desired deflection angle and wavelength input. The diversity of devices produced by the generator reflect the use of a random noise input in our probabilistic model. Details pertaining to the network structure and training process are in the Supplementary Section.

Our machine learning approach is qualitatively different from those based on feedforward neural networks, which use back-propagation for the inverse design of relatively simple nanophotonic devices *(23-27)*. These studies required tens of thousands of training data (*i.e.*, geometric layouts and their optical response) for the networks to learn the electromagnetic properties of shapes described by approximately ten geometric parameters. Complex shapes, on the other hand, are represented as images consisting of tens of thousands of pixels, described by hundreds of coefficients in the Fourier domain. Since the amount of required training data exponentially scales with the number of parameters describing the shapes *(28)*, the task of generating sufficient quantities of training data makes feedforward networks for complex shapes difficult to practically scale. With conditional GANs, we directly sample the space of high efficiency designs without the need to accurately predict the performance of every device along an optimization trajectory. The algorithms focus on learning important topological features harvested from high-performance metasurfaces, rather than attempting to predict the behavior of every possible device, most of which are very far from optimal. In this manner, these networks produce high efficiency, topologically-intricate metasurfaces with substantially less training data.

To illustrate the ability for our conditional GAN to generate devices with operating parameters beyond those of the training set, we use our trained generator to produce 5000 different layouts of devices operating at a 70 degree deflection angle and a 1200 nm wavelength. The GAN can generate thousands of devices within seconds, making it possible to produce large datasets, even larger than even the entire training dataset, with low computational cost. We then calculate device efficiencies using a rigorous coupled-wave analysis solver *(29)*, and the distribution of efficiencies is plotted as a histogram in Figure 3A. The histogram of device efficiencies produced from the conditional GAN shows a broad distribution. Notably, there exist devices in the distribution with efficiencies over 60% and as high as 62%. The presence of these devices indicates that our conditional GAN is able to learn and generalize features from the metasurfaces in the training set. To be clear, such learning is possible because the devices within this parameter space share related underlying physics that translates to trends in the device shape layouts. For the GAN to effectively work in a different physical parameter space, such as devices with differing refractive indices, training data covering those parameters would be required.

We quantify these device metrics with multiple benchmarks. First, we characterize 5000 random binary patterns with feature sizes similar to those in our training set (Figure S4A). The efficiency histogram of these devices shows that the best device is only 30% efficient, indicating that randomly-generated patterns all exhibit poor efficiencies. Second, we evaluate and plot the deflection efficiencies of devices in the training set that have been geometrically stretched, such that they diffract 1200 nm light to 70 degrees. The efficiency histogram of these devices is also plotted in Figure 3A and displays a maximum efficiency of only 53%. An analysis of the stretched training set across the whole parameter space is shown in Figure S5. Third, we take the stretched devices in the training set and deform them with random elastic distortions *(35)* to produce a set of 5000 quasi-random patterns. The results are summarized in Figure S6 and indicate that the GAN still achieves better performance than the randomly deformed training set for the large majority of wavelength-angle pairs. These comparisons between the GAN-generated and training set devices indicate that the GAN is able to extrapolate geometric features beyond the training set, and can properly utilize white noise inputs to produce a diversity of physically relevant shapes (Figure S7).

The high efficiency devices produced by the conditional GAN can be further refined with iterative topology optimization. This additional refinement serves multiple purposes. First, it further improves the device efficiencies. Second, it incorporates robustness to fabrication imperfections into the metagrating designs, which makes experimentally fabricated devices more tolerant to processing defects *(20)*. Details pertaining the theoretical and experimental analysis of robustness have been covered in other studies *(35)*. Third, it enforces other experimental constraints, such as grid snapping or minimum feature size. Relatively few iterations of topology optimization are required at this stage because the devices from the conditional GAN are already highly efficient and near a local optimum in the design space.

With this approach, we apply 30 iterations of adjoint-based topology optimization to the 50 highest efficiency GAN-generated devices from Figure 3A. With topology refinement, devices are optimized to be robust to geometric erosion and dilation. The final device efficiency distributions are plotted in Figure 3B. Interestingly, some of the devices have efficiencies that lower after topology optimization. The reason is that these devices from the generator were not initially robust, and their efficiencies were penalized as the optimizer enforced robustness constraints into the designs. The highest performance device has an efficiency of 86%, which is comparable to the best device from a distribution of iterative-only optimized devices (Figure S1). A plot of device efficiency over the course of iterative optimization for a representative metagrating is shown in Figure 3C. We note that for topology refinement, more iterations can be performed to further improve the devices, at the expense of computation cost. We consider 30 iterations of topology refinement here to balance computation time with final device quality.

Our strategy to design robust, high-efficiency metagratings with the GAN generator and iterative optimizer can apply to a broad range of desired deflection angles and wavelengths. With the same training data from before, we design robust metagratings with operating wavelengths ranging from 500 nm and 1300 nm, in increments of 50 nm, and angles ranging from 35 and 85 degrees, in increments of 5 degrees. 5000 devices are initially generated and characterized for each angle and wavelength, and topology refinement is performed on the 50 most efficient devices. Figure 4A shows the device efficiencies from the generator, where the efficiencies of the highest performing devices for a given angle and wavelength are presented. Most of the generated devices have efficiencies over 65%, and within and near the parameter space specified by the training set (green box), the generated devices have efficiencies over 75%.

Representative images of high efficiency metagratings from the generator are shown in Figure 4B. We find that at shorter wavelengths, the metagratings generally comprise spatially distributed dielectric features. As the wavelengths get longer, the devices exhibit more consolidated distributions of dielectric material with fewer voids. These variations in topology are qualitatively similar to those featured in the training set (Figure 1B). Furthermore, these trends in topology clearly extend to devices operating at wavelengths beyond those used in the training set. Additional images of devices from GAN generator are shown in Figure S8.

The device efficiencies of the best devices after topology refinement are presented in Figure 4C. We see that nearly all the metagratings with wavelengths in the 600–1300 nm range and angles in the 35–75 degree range have efficiencies near or over 80%. Not all the devices produced with our method exhibit high efficiencies, as Figure 4C shows clear drop-offs in efficiencies for devices designed for shorter wavelengths and ultra-large deflection angles. One source for this observed drop-off is that these devices are in a parameter space that requires topologically distinctive features that could not be generalized from the training set. As such, the conditional GAN is unable to learn the proper patterns required to generate high performance devices. There are also device operating regimes for which high efficiency beam deflection is not physically possible with 325 nm-thick silicon metagratings. For example, device efficiency will drop off as the operating wavelength becomes substantially larger than the device thickness *(30)* and when the deflection angle becomes exceedingly large (Figure S1, fifth column).

The capabilities of our conditional GAN can be enhanced by network retraining with additional data. These data can originate from two sources. The first is from iterative-only optimization, which is how we produced our initial metagrating training set. The second is from the GAN generator and topology refinement process. This second source of training data suggests a pathway to expanding the efficacy of our conditional GAN with high computational efficiency.

As a proof-of-concept, we use the generator and iterative optimizer to produce 6000 additional high efficiency (70%+) robust metagratings with wavelengths and angles spanning the full parameter space featured in Figure 4A. We then add these data to our previous training set and retrain our conditional GAN, producing a "second generation" GAN. Figure 5A shows the device efficiencies from the retrained generator, where 5000 devices for a given angle and wavelength are generated and the efficiencies of the highest performing devices are presented. The plot shows that the efficiency values of devices produced by the retrained GAN generally increase in comparison to those produced by the original GAN. Quantitatively, over 80% of the devices in the parameter space have improved efficiencies after retraining (Figure S9A).

The efficiency histograms of devices generated from our second generation GAN and then topology-refined are plotted in Figure 5B for representative wavelength and deflection angle combinations. For this topology-refinement step, 50 iterations of iterative optimization are performed for each device. The histograms of iterative-only optimized devices are also plotted as a reference. A more complete dataset including other device parameters is presented in Figure S1. These data indicate that for many wavelengths and deflection angles, the best topology-refined devices are comparable with the best iteratively-optimized devices: for 80% of the histograms in Figure S1, the best topology-refined device in a histogram has an efficiency that exceeds or is within 5% the efficiency of the best iteratively-optimized device. For 25% of the histograms in Figure S1, the best topology-refined device in a histogram has an efficiency that exceeds the efficiency of the best iteratively-optimized device.

At short operating wavelengths, the neural network approach produces efficiency histograms similar to those of the iterative-only optimized devices. However, for devices operating at small deflection angles and long wavelengths, our GAN-based approach does not compare as well with iterative-only optimization. There is much room for further improvement. First, the architecture of the neural network can be further optimized. For example, a deeper neural network such as ResNet *(31, 32)* can be used, and the network can be trained dynamically as the resolution of generated patterns progressively grows *(33)*. Second, the choice of parameters for the training dataset can be more strategically chosen and optimized. Third, there may be ways to incorporate physics and electromagnetics domain knowledge into the GAN. Fourth, we generally expect our GAN-based approach to improve as the training sets get larger.

Using our fully-trained second generation GAN, we estimate the computational time required to generate and refine "above threshold" devices, previously defined in our vetting of the training set earlier, across the full parameter space featured in Figure 4A. The results are summarized in Figure 5C and Table S2. We also include a trend line for devices designed using iterative-only optimization (red line). We find that the computational cost of designing "above threshold" devices using GAN generation, evaluation, and device refinement is relatively low. The result produces a trend for computational cost described by the blue line, which has a slope approximately five times less steep than that of the red line. The data used for this analysis are taken from the wavelength and angle pairs from Figure S1.

This analysis of the computation cost for generating and refining metagratings indicates that our trained GAN-based generator can be utilized as a computationally-efficient design tool. Consider a device parameter space that is of general interest for device design. Prior to designing any devices, we can first produce a set of devices within this parameter space and train our conditional GAN. The computational resources here can be treated as a onetime "offline" cost. Then, when a set of devices is desired, we can utilize our GAN-based approach to design the devices. With the computational cost of the training data already paid, our approach to device design will be faster than iterative-only optimization, as indicated by the relative slopes in Figure 5C.

**Conclusions**

In summary, we show that generative neural networks can facilitate the computationally efficient design of high performance, topologically-complex metasurfaces in cases where it is of interest to generate a large family of designs. Neural networks are a powerful and appropriate tool for this design problem because there exists a strong interdependence between device topology and optical response, particularly for high performance devices. In addition, we have the capability to generate high quality training data and validate device performance using the combination of iterative optimizers and accurate electromagnetic solvers.

While this study focuses on the variation of two device parameters (*i.e.*, wavelength and deflection angle), one can imagine generalizing the GAN-based approach to more device parameters, such as device thickness, device dielectric, polarization, phase response, and incidence angle. Such multifunctional devices can potentially be realized using a high quality dataset of multifunctional devices or by implementing multiple discriminators for pattern synthesis. Iterative-only optimization methods simply cannot scale to these high dimensional parameter spaces, making data-driven methods a necessary route to the design of large numbers of topologically-complex devices. We also note that generative networks can be directly integrated in the topology optimization process, by replacing the discriminator with an electromagnetics simulator *(36)*. In all of these embodiments of generative networks, the efficient generation of large datasets of topologically-optimal metasurfaces paves the way for the use of other machine learning and data mining schemes for device analysis and generation.

We envision that data-driven design processes will apply to the design and characterization of other complex nanophotonic devices, ranging from dielectric and plasmonic antennas to photonic crystals. The methods we described can also encompass the design of devices and structured materials in other fields, such as acoustics, mechanics, and heat transfer, where there is a need to design functional elements across a broad parameter space.

**Methods**

The network architectures of the conditional GAN are shown in Table S1. The input to the generator is a 128x1 vector of Gaussian random variables, the operating wavelength, and the output deflection angle. All of these input values are normalized to numbers between -1 and 1. The output of the generator, as well as the input to the discriminator, are binary images on a 64x256 grid, which is half of one unit cell. Mirror symmetry along the y-axis is enforced by using reflecting padding in the convolution and deconvolution layers. Periodic padding is also used to capture the periodic nature of the metagratings. We also include multiple copies of the same devices in the training set, with each copy randomly translated along the x-axis.

We find that the GAN generator tends to create slightly noisy patterns with very small features. These features are not present in devices in the training set, which are robust to fabrication errors and minimally utilize small feature sizes. To generate devices that better mimic those from the training dataset, we add a Gaussian filter at the end of the generator, before the tanh layer, to eliminate any fine features in the generated devices.

During the training process, both the generator and discriminator use the Adam optimizer with a batch size of 128, learning rate of 0.001, beta1 of 0, and beta2 of 0.99. We use the improved Wasserstein loss *(37, 38)* with a gradient penalty, with lambda = 10. The network is implemented using Tensorflow and trained on one Tesla K80 GPU for 1000 iterations, which takes about 5-10 minutes.

**Acknowledgments:** The simulations were performed in the Sherlock computing cluster at Stanford University. **Funding:** This work was supported by the U.S. Air Force under Award Number FA9550-18-1-0070, the Office of Naval Research under Award Number N00014-16-1-2630, and the David and Lucile Packard Foundation. DS was supported by the National Science Foundation (NSF) through the NSF Graduate Research Fellowship. **Competing interests:** Authors declare no competing interests.

30. Yang J, Fan JA (2017) Analysis of material selection on dielectric metasurface performance. Opt. Express 25, 23899-23909.
31. He, K., Zhang, X., Ren, S. and Sun, J., 2016. Deep residual learning for image recognition. In Proceedings of the IEEE conference on computer vision and pattern recognition (pp. 770-778).
32. Brock A, Donahue J, Simonyan K. Large scale gan training for high fidelity natural image synthesis. arXiv preprint arXiv:1809.11096. 2018 Sep 28.
33. Karras T, Aila T, Laine S, Lehtinen J. Progressive growing of gans for improved quality, stability, and variation. arXiv preprint arXiv:1710.10196. 2017 Oct 27.
34. Phan T, Sell D, Wang EW, Doshay S, Edee K, Yang J, Fan JA. High-efficiency, large-area, topology-optimized metasurfaces. Light: Science & Applications. 2019 May 29;8(1):48.
35. Wang EW, Sell D, Phan T, Fan JA. Robust design of topology-optimized metasurfaces. Optical Materials Express. 2019 Feb 1;9(2):469-82.
36. Jiang J, Fan JA. Global optimization of dielectric metasurfaces using a physics-driven neural network. arXiv preprint arXiv:1906.04157. 2019 May 13.
37. Arjovsky M, Chintala S, Bottou L (2017) Wasserstein Generative Adversarial Networks. *International Conference on Machine Learning*, 214–223.
38. Gulrajani I, Ahmed F, Arjovsky M, Dumoulin V, Courville AC. Improved Training of Wasserstein GANs. *Advances in Neural Information Processing Systems* **2017**, 5767–5777.
39. Liu Z, Zhu D, Rodrigues SP, Lee KT and Cai W (2018) Generative model for the inverse design of metasurfaces. *Nano letters*, *18*(10), pp.6570-6576.
10

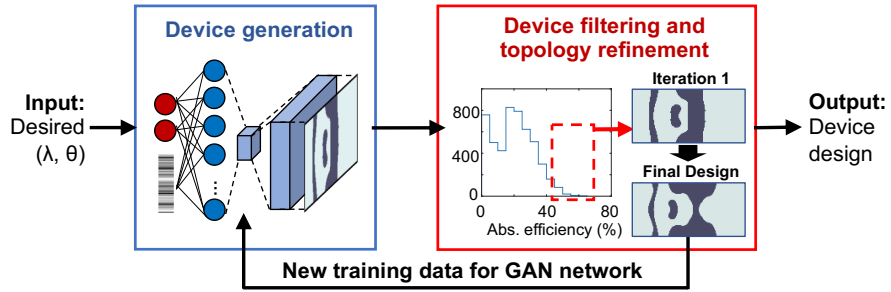

**Figure 1.** Schematic of metasurface inverse design based on device generation from a trained generative neural network, followed by topology optimization. Devices produced in this manner can be fed back into the neural network for retraining and network refinement.

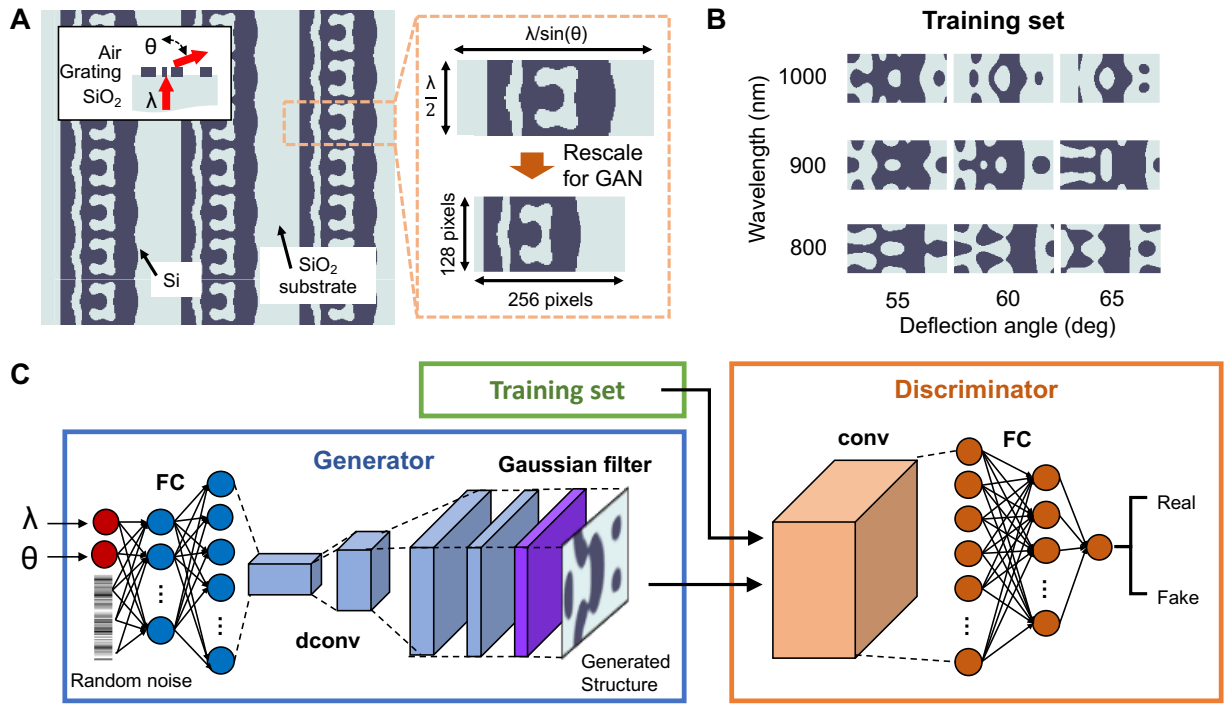

**Figure 2.** Machine learning with topology-optimized metagratings. (A) Top view image of a typical topology-optimized metagrating that selectively deflects light to the +1 diffraction order. The metagratings are made of 325nm-thick Si layer on top of a $SiO_2$ substrate. The input data to the GAN are images of single metagrating unit cells rescaled to a 128 x 256 pixel grid. (B) Representative images of metagratings in the training set. All devices deflect TE-polarized light with over 75% efficiency, and each is designed to operate for a specific deflection angle and wavelength. (C) Schematic of the conditional GAN for metagrating generation. The generator utilizes two fully connected (FC) and four deconvolution (dconv) layers, followed by a Gaussian filtering layer, while the discriminator utilizes one convolutional (conv) layer and two fully connected layers. After training, the generator produces images of new, topologically-complex devices designed for a desired deflection angle and wavelength.

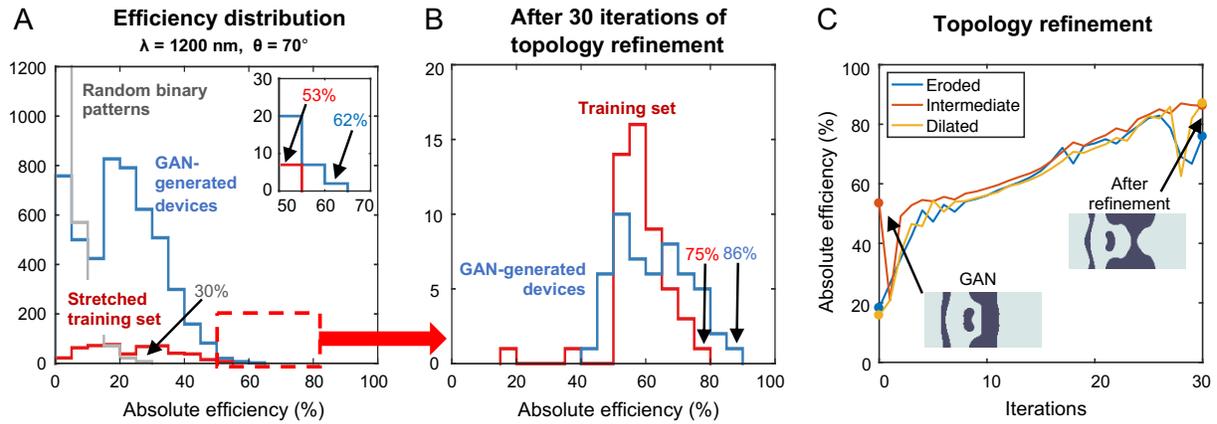

**Figure 3.** Metagrating generation and refinement. (A) Efficiency histograms of metagratings produced from the trained GAN generator, training set geometrically stretched for the design wavelength and angle, and randomly generated binary patterns. The design wavelength and angle for these devices are 1200 nm and 70 degrees, respectively, which is beyond training set. The highest device efficiencies in the histograms are displayed. Inset: Magnified view of the histogram outlined by the dashed red box. (B) Efficiency histogram of metagratings from the trained GAN generator and training set, refined by topology optimization. The 50 highest efficiency devices from the GAN generator and training set are considered for topology refinement. The highest efficiency devices produced from the training set and GAN generator are displayed. (C) Efficiency of the eroded, dilated and intermediate devices as a function of iteration number. After topology refinement, device efficiency and robustness are improved. Inset: Top view of the metagrating unit cell before and after topology refinement.
13

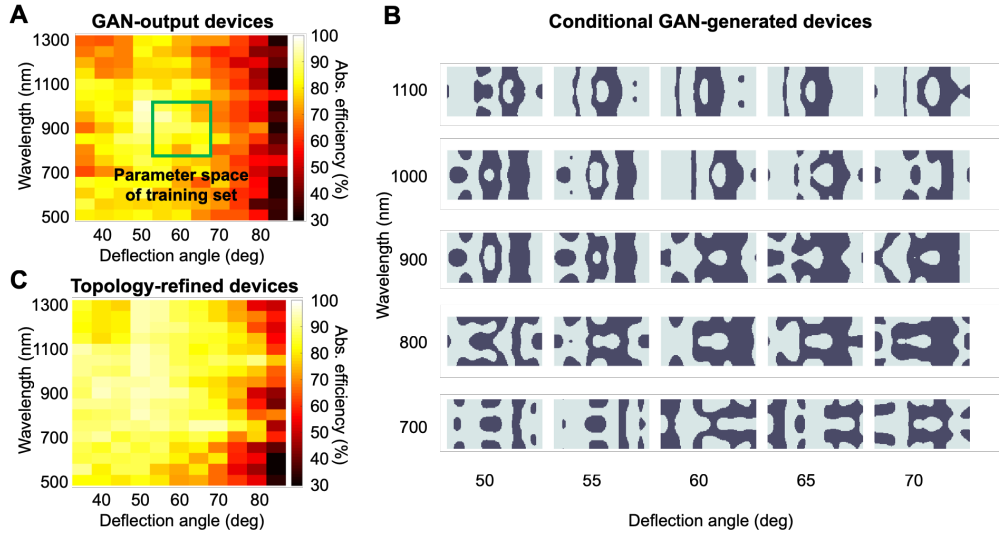

**Figure 4.** Metagrating performance across a broad parameter space. (A) Plot of the highest device efficiencies for metagratings produced by the GAN generator for differing wavelength and deflection angle parameters. The solid yellow box represents the range of parameters covered by devices in the training set. (B) Representative images of high efficiency metagratings produced by the GAN generator for differing operating wavelengths and angles. (C) Plot of the highest device efficiencies for metagratings generated from the conditional GAN and then topology-optimized.

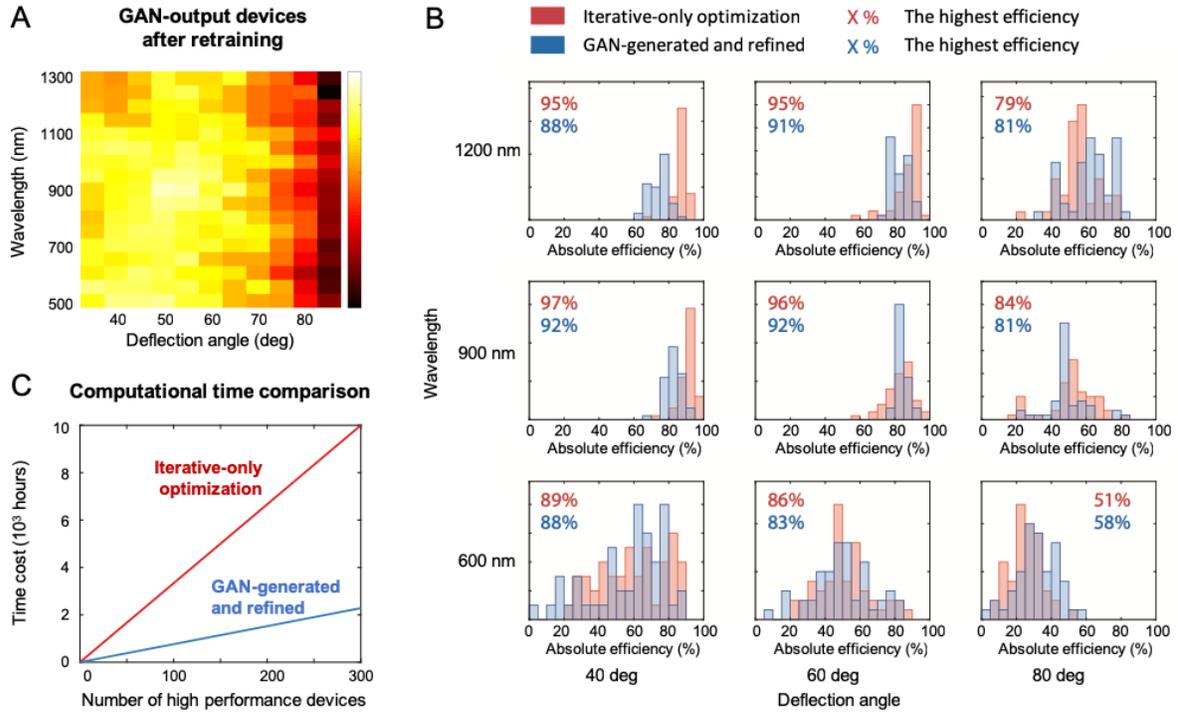

**Figure 5.** Benchmarking of GAN-based computational cost and network retraining efficiency. (A) Plot of the highest device efficiencies for metagratings produced by a retrained GAN generator. The initial training set is supplemented with an additional 6000 high efficiency devices operating across the entire parameter space. (B) Representative efficiency distributions of devices designed using iterative-only optimization (red histograms) and generation of the retrained GAN with topology refinement (blue histograms). The highest efficiencies are denoted by red numbers and blue numbers. (C) Time cost of generating $n$ "above threshold" devices using iterative-only optimization (red line) and GAN-generation and refinement (blue line). "Above threshold" is defined in the main text.

15